%% file: main.tex
\documentclass[conf]{new-aiaa}

\usepackage[utf8]{inputenc}

\usepackage{graphicx}
\usepackage{amsmath}
\usepackage[version=4]{mhchem}
\usepackage{siunitx}
\usepackage{longtable,tabularx}
\usepackage{placeins}
\usepackage{booktabs}
\usepackage{longtable}

\usepackage{booktabs}

\title{Human-in-the-Loop Testing of AI Agents for Air Traffic Control with a Regulated Assessment Framework}

\author{Ben Carvell\footnote{Co-Investigator, Project Bluebird, NATS} and Marc Thomas\footnote{Co-Investigator, Project Bluebird, NATS}}
\affil{NATS, Whiteley, Hampshire, PO15 7FL, United Kingdom}
\author{Andrew Pace\footnote{Operational ATCO and Group Supervisor, London Terminal Control, NATS} and Christopher Dorney\footnote{Initial Training Manager, NATS}}
\affil{NATS, Whiteley, Hampshire, PO15 7FL, United Kingdom}
\author{George {De~Ath}\footnote{Lecturer, Department of Computer Science, University of Exeter}}
\author{Richard Everson\footnote{Principal Investigator, Project Bluebird, University of Exeter and The Alan Turing Institute}
}
\affil{University of Exeter, Exeter, England EX4 4QJ, United Kingdom}
\author{Nick Pepper\footnote{Co-Investigator, Project Bluebird, The Alan Turing Institute} and Adam Keane\footnote{Research Associate, Project Bluebird, The Alan Turing Institute} }
\affil{The Alan Turing Institute, London, England NW1 2DB, United Kingdom}
\author{Samuel Tomlinson\footnote{Senior Researcher, Project Bluebird, NATS} and Richard Cannon\footnote{Principal Investigator, Project Bluebird, NATS} }
\affil{NATS, Whiteley, Hampshire, PO15 7FL, United Kingdom}

\begin{document}

\maketitle

\begin{abstract}
We present a rigorous, human-in-the-loop evaluation framework for assessing the performance of AI agents on the task of Air Traffic Control, grounded in a regulator-certified simulator-based curriculum used for training and testing real-world trainee controllers. By leveraging legally regulated assessments and involving expert human instructors in the evaluation process, our framework enables a more authentic and domain-accurate measurement of AI performance. This work addresses a critical gap in the existing literature: the frequent misalignment between academic representations of Air Traffic Control and the complexities of the actual operational environment. It also lays the foundations for effective future human-machine teaming paradigms by aligning machine performance with human assessment targets.
\end{abstract}

\section{Introduction}

\lettrine{A}{ir} Traffic Control (ATC) is a complex, safety-critical decision-making task that is executed in the presence of significant epistemic uncertainty~\citep{lymperopoulos2010sequential}. Air Traffic Control Officers (ATCOs) issue mandatory instructions to aircraft to maintain and ensure safety between them while facilitating efficient flight trajectories that conform to all relevant sector procedures~\citep{CAP493}. While a broad array of support systems are currently used in operations around the world, no systems exist that automate the decision-making process for controllers. ATC remains a manual task, with tools identifying potential conflicts between aircraft and providing predictions of their flight paths, without offering suggestions of what actions to take.

\subsection{Prior Work: ATC Automation Research}
The automation of this task has been the subject of industrial and academic research for more than five decades. Earlier efforts utilised rules-based heuristic systems to solve conflicts and guide aircraft \citep{wesson_planning_1977, ARC2000,erzberger_automated_2006}, and over recent years, the leading edge of artificial intelligence techniques such as deep reinforcement learning \citep{brittain_autonomous_2019,sui_conflict_2023,wang_deep_2019} and a range of multi-agent methods \citep{pappas_conflict_1996,tumer_improving_2009} have been applied with the hope of progressing towards real domain applications for decision automation.

Despite half a century of research, any decision automation in ATC still seems distant. The closest a system has come is the ARGOS prototype system under development by Eurocontrol~\citep{argos}, but this system is not yet in operation, and the solution methods are built on heuristics rather than state-of-the-art AI techniques.

We argue that a major contributing factor to this lack of progress is rooted in the fundamental misalignment between the representation of the ATC problem in leading-edge academic research and the realities of ATC operations. Simplifications are made to ensure compatibility with the chosen technology; thus, true domain relevance remains out of reach for powerful machine learning-based methods. This is a blocker for future concepts of more involved human-machine collaboration, as the tasks being executed are fundamentally not the same.

\subsection{Prior Work: Human-in-the-Loop and Curriculum-Driven Evaluation}

The evaluation of safety-critical autonomy requires methodologies that transcend simple numerical metrics. \citet{crootof_humans_2022} emphasise that regulatory frameworks necessitate Human-in-the-Loop (HITL) verification to account for the complexities of the real world in high-stakes automation. \citet{wu_survey_2022} describe how the involvement of humans in the process of machine learning can significantly improve performance through embedding expert knowledge in labelling data, domain-driven reward engineering, and incorporating human experience in the modelling process. To better align with human-centric measures of performance, researchers have examined the use of established professional licensing curricula as evaluation benchmarks. While recent studies have successfully tested large language models against text-based standards, such as the US Medical Licensing Exam~\citep{kung_performance_2023} and the Uniform Bar Examination (UBE)~\citep{katz_gpt-4_nodate}, physical control tasks require more dynamic assessment environments. This need is addressed in the autonomous driving domain through scenario-based safety assessment. The work of \citet{riedmaier_survey_2020} details how these methodologies decompose the operational domain into functional scenarios, offering a structured approach to safety assurance that complements formal verification methods.

\subsection{Motivation and Contribution}

\begin{figure} 
    \centering
    \includegraphics[width=1.0\textwidth]{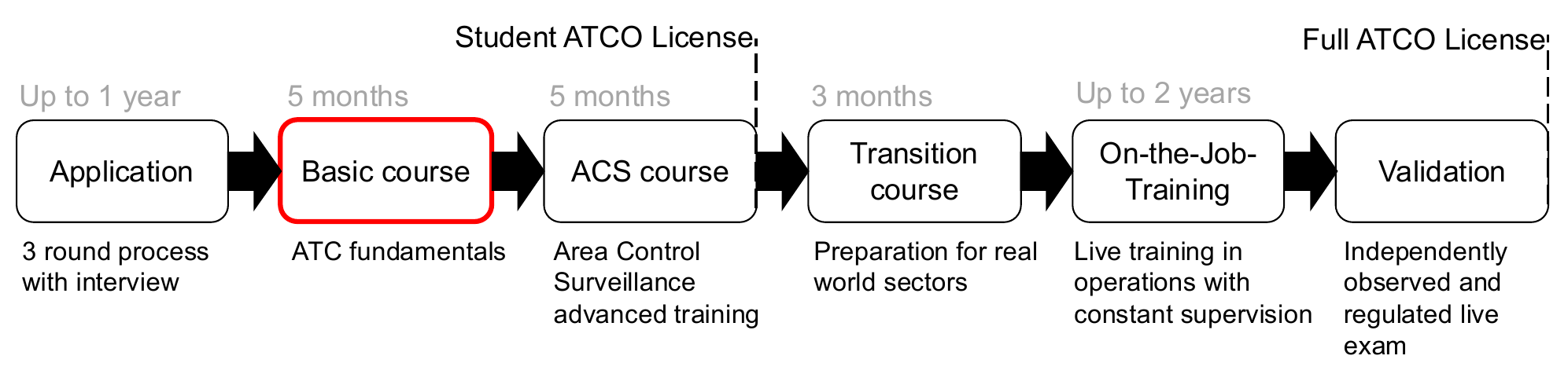}
    \caption{The progression of training to become a licensed ATCO at NATS \cite{NATS_ATCO_pathway}.}
    \label{fig:training_progress} 
\end{figure}

In this paper, we demonstrate a framework to address the misalignment of research representations of ATC by leveraging the \textit{Basic} training course provided by NATS to new trainee ATCOs. Through \textbf{\textit{Project Bluebird}}\footnote{Project Bluebird is an EPSRC Prosperity Partnership between NATS, The Alan Turing Institute and The University of Exeter.}, a probabilistic digital twin of the UK ATC environment has been created, which we refer to as BluebirdDT~\citep{bluebird_dt}. The configurable nature of this system allows for a range of airspace, aircraft performance models, and action spaces to be simulated. With this digital twin as a base environment, we have adapted the assessment framework used by NATS to train human ATCOs, in order to allow it to be used by machines. We do not use the full operational definition of ATC, but the fundamentals taught in the NATS ATC college Basic course, shown in context in Figure~\ref{fig:training_progress}. This course is driven by core regulatory requirements from the UK ATC regulator, the Civil Aviation Authority (CAA), and provides a framework that is less complex than the challenge of full operational ATC but captures the core elements of the task, unlike most academic simplifications.

Through human-in-the-loop assessment, we can at last give
more authentic and domain-aligned measures of progress that are comparable across a wide range of automation techniques.
We argue that the complexity of the ATC task and the difficulty in translating good controlling practise into quantitative metrics necessitate a human-in-the-loop approach that is distinct from the validation through scenario generation model that has been used extensively to validate automated driving technologies (see, e.g.,~\citep{riedmaier_survey_2020}).

We describe the competencies taught in the Basic course, the legal requirements which underpin these competencies, and our translation of this curriculum into a `Machine Basic Training' framework. We discuss the results of trials utilising the framework for agent assessment and set out future plans for using the framework to build domain-accurate measures of success for Air Traffic Control, transforming the ability of researchers to effectively engage with this highly challenging domain.

\section{Methodology: Machine Basic Training (MBT)}

Through a year-long series of workshops with operational ATCOs and instructors, we have designed a HITL framework for agent training and assessment, which we term Machine Basic Training (MBT). This framework aims to adapt the curriculum from the initial ATC training course at NATS for an agent-based approach, recreating the training and assessment process to the highest fidelity possible while providing concrete guidance for performance. In this section, we will summarise the Basic course, which is the basis of MBT, the legal and regulatory foundations of this course, the agent development and assessment scope that we have created to adapt this curriculum for machines, and the procedure used to test agents within this framework.

\subsection{The `Area Basic' Training Course}\label{sec:basic}

The `Area Basic' training course is a 5-month programme taught at NATS' training college to all new trainee ATCOs. It consists of a comprehensive set of classroom-based training on ATC concepts, such as legal considerations, weather, aircraft performance, controlling techniques, radio-telephony discipline, and more. Alongside this classroom-based study, students control a series of approximately 30 simulator exercises with guidance from an instructor. These exercises take place on a high-fidelity ATC simulator, with a `pseudo-pilot' playing the role of the aircraft pilots within the exercise. Instructions are issued via voice communication to replicate radio communication in the real world. A fictional airspace sector, `Medway', is used for these exercises. The Medway sector is designed to accurately reflect a range of en route environments and is structured to provide challenging scenarios. Students use a radar to monitor the ATC situation and paper flight strips to record instructions given, detect potential conflicts between aircraft, and to interpret aircraft characteristics.

Each of these `formative' exercises is graded to track the student's progress against a set of 6 competency areas, as shown in Table \ref{tab:competency_areas}.

\begin{table}[h]
\centering
\caption{Competency Areas Assessed on Basic Exercises.}
\begin{tabular}{p{2.5cm}|p{12cm} } 
 Competency & Description\\ [0.5ex] 
 \hline
 Safety & Safety is the core competency for ATC. In the Basic course Aircraft must be kept apart by a minimum of 5 nautical miles laterally or 1000 feet vertically. However, safe controlling encompasses much more than these minimum standards. A method of controlling is taught which not only achieves these standards, but attempts to ensure that they will always be met if no further instructions are issued. This fail-safe method of operation underpins all the taught controlling technique. An example of safe controlling technique is shown in Figure~\ref{fig:safety_demo}\\ 

 Controlling & Controlling assesses a student's ability to effectively instruct the aircraft under their control to achieve sector goals such as maintaining and ensuring safety, achieving agreed exit coordinations, and remaining within the confines of the sector airspace. Controlling techniques to achieve this include accurate vectoring, timely climb and descent instructions, and appreciation of aircraft performance. An example of this is shown in Figure~\ref{fig:controlling_demo}.\\ 

 Planning & Planning covers a student's ability to create effective plans for managing the sector. Proactive planning which takes into account future traffic patterns and interactions will allow for more efficient control, and is necessary to enable achievement of sector goals.\\
 
 Coordination & Coordination is the process of making and fulfilling agreements with neighbouring sectors regarding the presentation and transfer of traffic between them, specifying a point of transfer and a level for the aircraft to be transferred. All aircraft must be coordinated into and out of the sector, with these agreements made in a timely manner and obeyed.\\

 Communication & Communication tracks the ability of the student to effectively communicate with aircraft. It includes the confident delivery of correct phraseology as described in the CAP413 Radio-telephony manual~\citep{CAP413}, as well as effective management of the sector frequency, responding to pilot requests, and combining instructions where appropriate.\\

Human Factors &  Human factors covers the student's attitude to training, resilience in the face of mistakes, and various more human-centric characteristics which may not fit into the other described competencies.\\
 [1ex]
\end{tabular}

\label{tab:competency_areas}
\end{table}

\begin{figure} [h]
    \centering
    \includegraphics[width=0.8\textwidth]{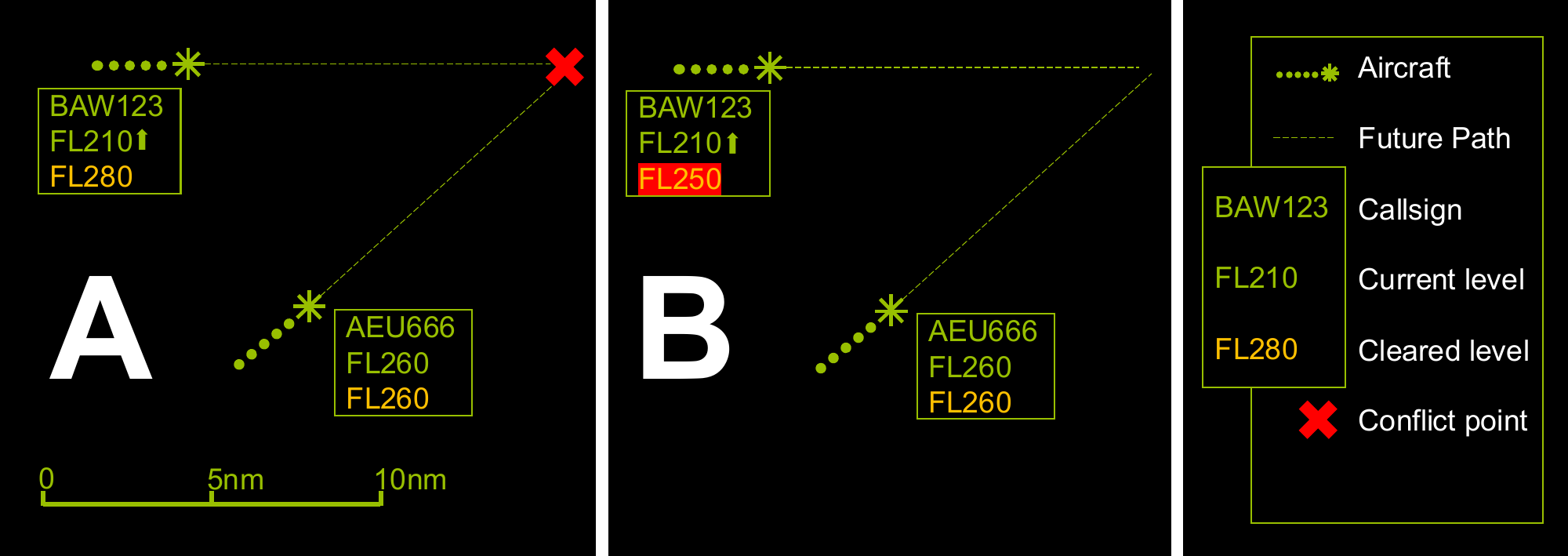}
    \caption{Judgement of safety in plan view. In scenario A (left) BAW123 is in conflict with AEU666, with no lateral or vertical separation ensured, meaning that a risk of future collision exists. In scenario B (middle) BAW123 has been climbed to a level 10 flight levels (1000 feet) below AEU666, ensuring safety between them.}
    \label{fig:safety_demo} 
\end{figure}

\begin{figure} [h]
    \centering
    \includegraphics[width=0.8\textwidth]{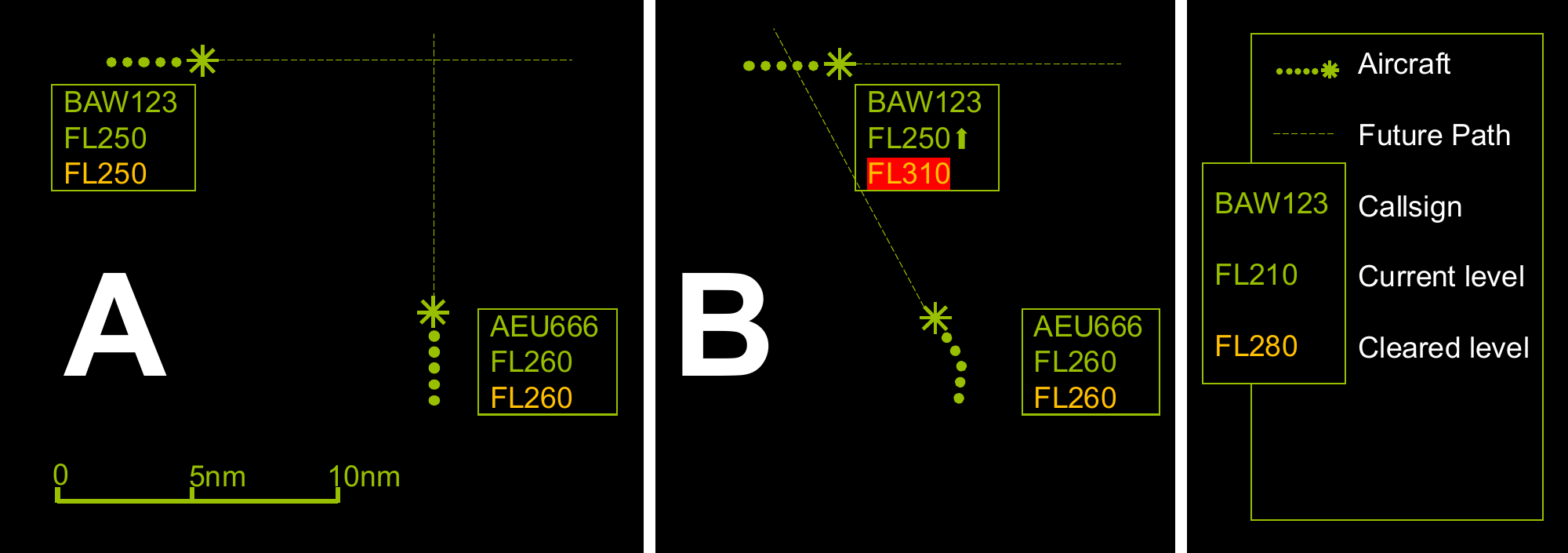}
    \caption{Judgement of effective controlling in plan view. In scenario A (left) BAW123 is unable to climb due to a potential conflict with AEU666. In scenario B (middle) AEU666 has been turned behind such that BAW123 can climb without risk of collision.}
    \label{fig:controlling_demo} 
\end{figure}

\subsection{Legal Foundations of Training}\label{sec:legal}
\begin{figure} [!h]
    \centering
    \includegraphics[width=1.0\textwidth]{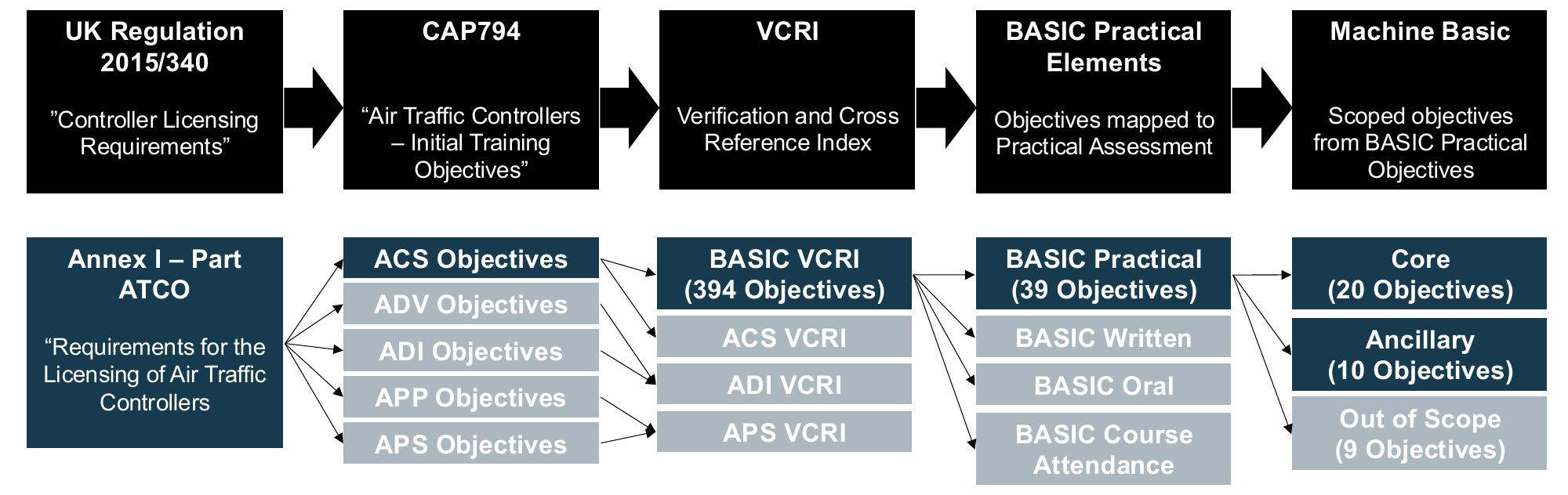}
    \caption{The flow down of requirements from top-level CAA regulations to machine basic.}
    \label{fig:flowdown} 
\end{figure}

As a regulated industry, NATS has a set of legal requirements that it must fulfil when providing training, and these are processed through a set of documents, as shown in Figure~\ref{fig:flowdown}. UK Regulation 2015/340~\citep{CAAreg2015} comprises the full legal requirements for training, which are organised and given additional detail in CAP794~\citep{CAP794}, grouping objectives from the original UK Regulation into sets depending on ATC discipline. The Basic course focusses on Area Control, and these requirements are mapped in an internal document, the Verification and Cross Reference Index which describes coverage against the regulation objectives. Within this document, objectives are assigned to tuition and assessment methods. For the purposes of MBT, we consider only those objectives that are measurable through practical assessment, comprising 39 objectives. Within these, we may then define the scope we are able to assess. These 39 objectives broadly cover the 6 competency areas discussed in Section~\ref{sec:basic}. Other objectives that are not covered in practical assessment comprise subjects such as aviation law, understanding ATC systems, and other knowledge-based elements that are assessed through written and oral exams; therefore, they fall outside the scope of this framework. Further details on scoping can be found in Section~\ref{sec:scoping}.

\subsection{Scope for Agent Development and Assessment}\label{sec:scoping}
Effective assessment of agents against the Basic curriculum relies on clear definitions of which behaviours are in and out of scope. As discussed in Section~\ref{sec:legal}, the NATS internal Verification and Cross Reference Index (VCRI) describes a set of 39 objectives that are measured through practical assessment. These objectives are formally mapped to regulatory obligations, as shown in Figure~\ref{fig:flowdown}, and take the form of elements relating to the six core competency areas in Table~\ref{tab:competency_areas}. They take the form of elements of these competencies, such as \textit{``Identify separation problems''} or \textit{``Appreciate relative velocity between aircraft''}. In summary, we focus on Safety, Controlling, Planning, and Coordination-related objectives and currently place Communication and Human Factors out of scope. We still emulate effective communication, but this is facilitated through the simulator API rather than agent competency in phraseology and radio-telephony. Translation of human factors considerations for the machine will form a critical part of future work. Objectives where we make a partial assessment are marked as `Ancillary', such as coordination, where we fix target conditions for aircraft while still monitoring for agents successfully achieving those conditions.

The final stage of scoping we have performed is the translation of these internal NATS training objectives into more explicit and fine-grained descriptions of which behaviours are in and out of scope, heavily drawing on the ICAO ``Manual on Air Traffic Controller Competency-based Training and Assessment''~\cite{icao10056} as a source for more granular descriptions of the controlling task. The selection of objectives was decided and refined through a series of 15 workshops with both operational ATCOs and current instructor staff. The aim of these workshops was to retain as broad a scope as possible whilst removing objectives which add engineering complication to agent design without forming part of the core tactical deconfliction and control task. The set of objectives derived from this exercise forms the fundamental scope for MBT and is detailed in Section \ref{sec:mbt_appendix}.

\subsection{Formative and Summative Exercises}
The Area Basic course provides a series of formative (training) and summative (evaluation) exercises, which we have recreated within the BluebirdDT Digital Twin platform. The formative exercises comprise around 30 scenarios of increasing difficulty, which agents can access and run at will. These ``formatives'' have been the foundation of many agent design workshops within Project Bluebird and provide a method for agents to demonstrate their behaviour and be judged by instructors, similar to training in the real ATC college. The summative exercises are held back as a hidden test set, in the same way that real trainees are not exposed to these exercises prior to examination.
Due to proprietary intellectual property constraints, the formative and summative exercises utilised in this work cannot be publicly disclosed. However, as detailed further in Section \ref{sec:future_work}, we intend to release an equivalent set of open-source scenarios and associated airspace specifications to facilitate the utilisation of this framework.

\subsection{Assessment Procedure}

To run assessments with this framework, we utilise the `summative' exercises from the Basic course. We run a set of 3 simulations, each lasting thirty minutes, with an instructor observing agent performance. Instructors are briefed on the key elements of the MBT prior to the start of the run, including elements of the task that are marked out of scope by contrast to the standard Basic curriculum. Instructors observe the runs and write a report covering the competencies described in Section~\ref{sec:basic} for each run. Each competency is scored on a four-point scale that measures how well it has been achieved. The four grades are shown in Table \ref{competency_levels}.

\begin{table}[h!]
\centering
\caption{Competency achievement levels}
\begin{tabular}{p{0.25\linewidth} p{0.65\linewidth}}
\textbf{Level} & \textbf{Description} \\ \hline
Fully Achieved &
\textit{This competency has been fully achieved.} \\[1ex]
Mostly Achieved &
\textit{This competency has been mostly achieved with some minor errors.} \\[1ex]
Partly Achieved &
\textit{This competency has been only partially achieved, with significant errors and cause for concern.} \\[1ex]
Not Achieved &
\textit{This competency has not been achieved with major errors and/or a lack of technique observed.} \\
\end{tabular}
\label{competency_levels}
\end{table}

These reports are then judged together by a certified assessor to make the final assessment. There is no absolute standard as to what constitutes a passing grade within this marking scheme. A key feature of the assessment process is the writing of detailed report forms, which provide evidence and justification for the grading against each competency. As such, any deviation from \textit{Fully Achieved} will be individually examined by the assessor, who will consider all evidence provided by the set of summative assessment forms before coming to an overall conclusion. Small, isolated errors are unlikely to result in a failure, whereas clear patterns from evidence of repeated errors or flaws in technique will be identified and may result in an unsatisfactory rating. The final judgement for a candidate is provided as a \textit{Satisfactory} or \textit{Unsatisfactory} grading across each measured competency by the Assessor. A candidate must achieve \textit{Satisfactory} in all competency areas to achieve a pass.

\section{Candidate Agents}

This framework is agent-agnostic, and any approach can be used as long as it is able to interact with the simulator API to control traffic. To test this framework, we have run assessments of two prototype agents from Project Bluebird\cite{bluebird} that are still in development, allowing for useful feedback to be gathered for the agent developers and to stress-test the assessment framework. Both agents have the same simulation action API and have access to the same information. Further agents, such as the work of Kent et al.~\citep{bluebird_mallard} are in development and will be tested against this framework when mature.

\begin{figure}[h!]
\centering%
\begin{minipage}[t]{.5\textwidth}%
  \centering%
  \captionsetup{width=.8\linewidth} %
  \includegraphics[width=.9\linewidth,clip,trim={0 0 0 0}]{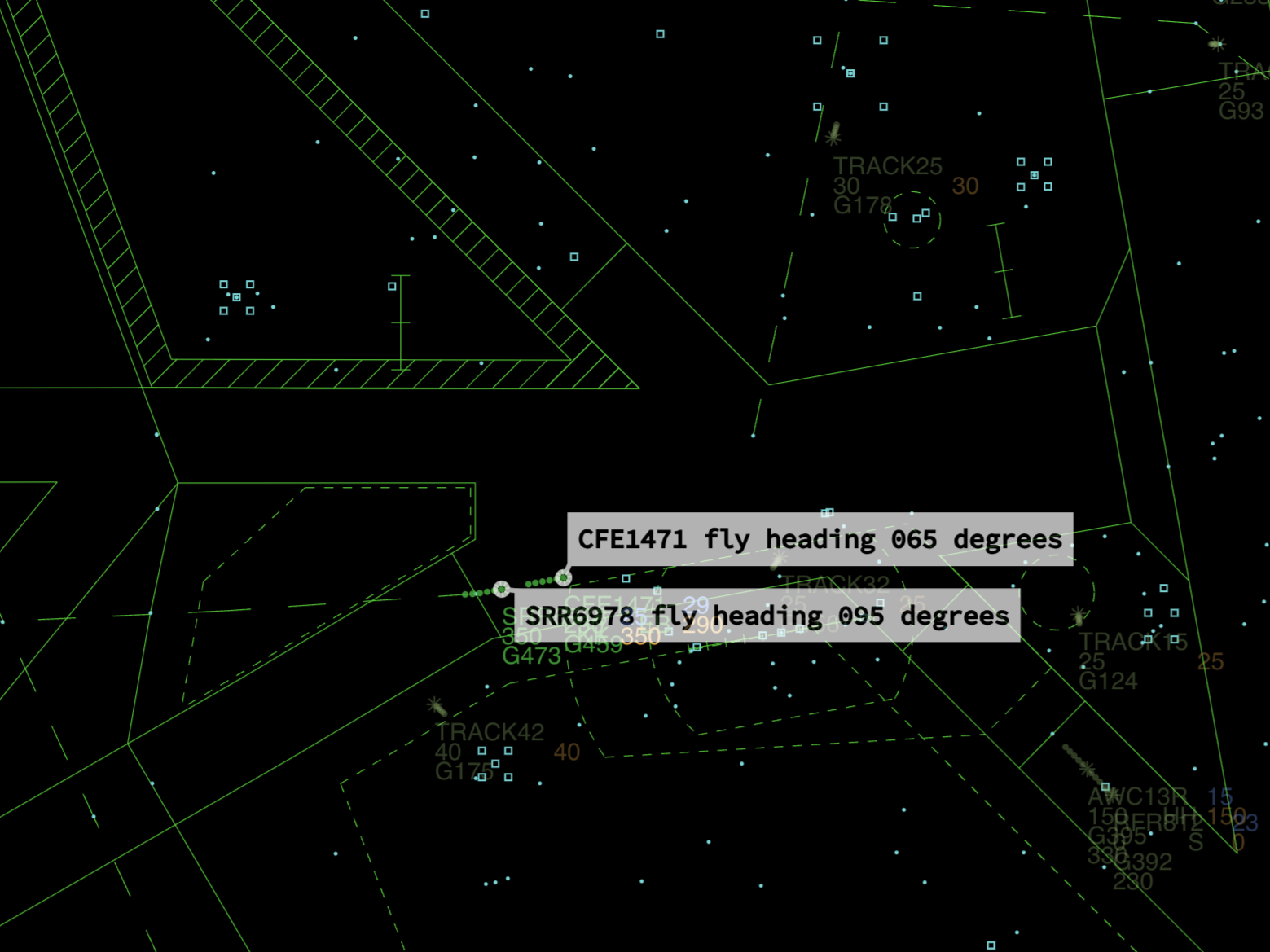}%
  \captionof{figure}{Rules-based agent \textit{Hawk} controlling an exercise in the BluebirdDT simulation environment.}
  \label{fig:hawk}%
\end{minipage}%
\begin{minipage}[t]{.5\textwidth}%
  \centering%
  \captionsetup{width=.8\linewidth}%
  \includegraphics[width=.9\linewidth]{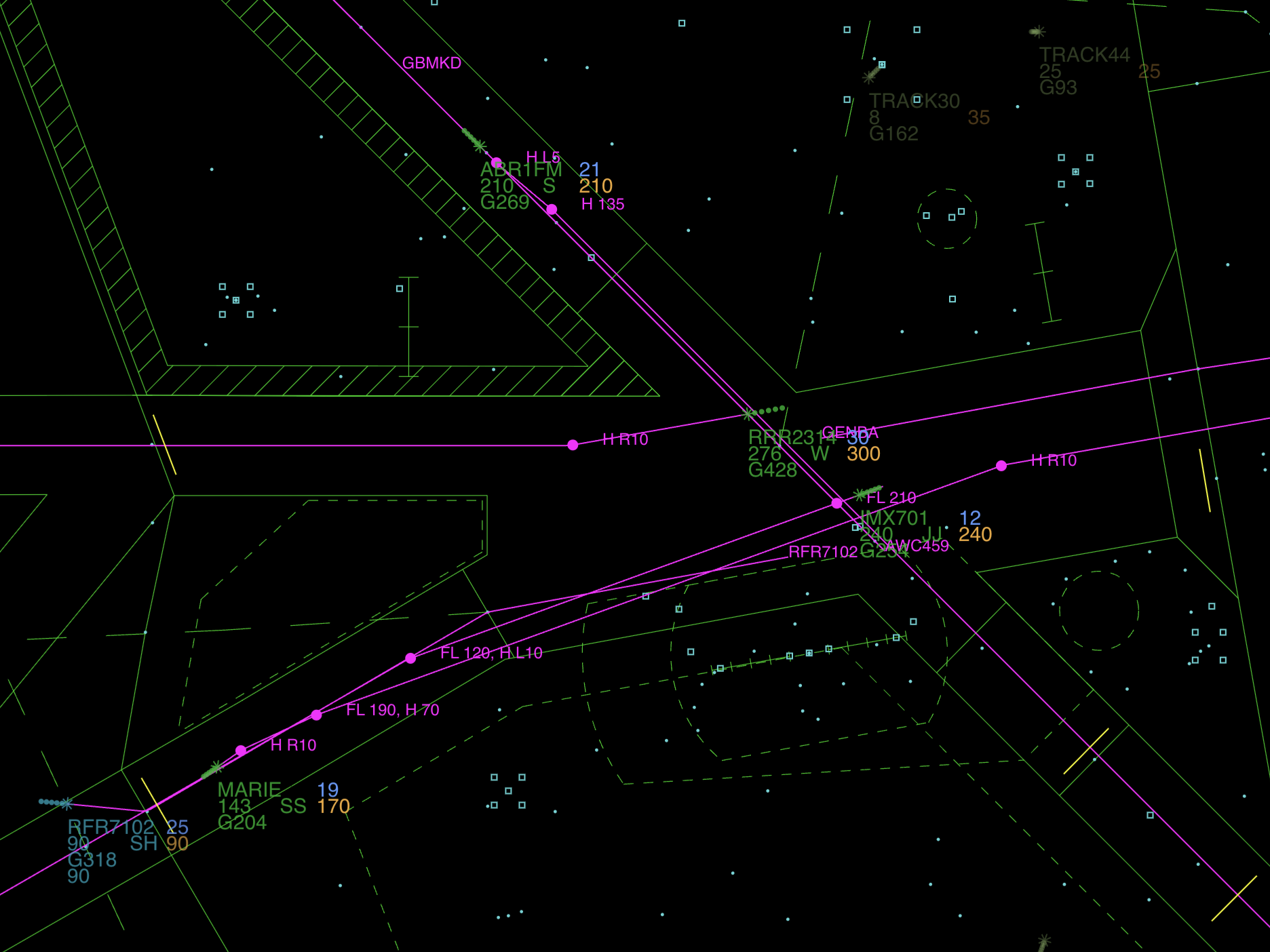}%
  \captionof{figure}{Optimisation agent \textit{Falcon} controlling an exercise in the BluebirdDT simulation environment. Planned trajectories are shown in magenta.}
  \label{fig:falcon}%
\end{minipage}
\end{figure}

\subsection{Rules-based Agent \textit{Hawk}}

Hawk is a rules-based agent built on rules created from extensive expert interviews, discussions with instructors on formative exercise performance, and interpretation of training materials. It utilises a decision tree to execute two primary rule sets: main rules, which deal with maintaining safety and conflict resolution, and iterative rules, which attempt to minimise path length and time flying below its cruising level flight. Figure~\ref{fig:hawk} shows Hawk controlling an exercise in the BluebirdDT simulation environment, where it separates two aircraft laterally by issuing two complementary heading instructions to ensure that the aircraft exchange flight levels safely.

\subsection{Optimisation-based Agent \textit{Falcon}}

Falcon is an optimisation-based agent. The planned trajectory of each aircraft through the sector is represented by a sequence of way-points and associated flight levels. These are adjusted by an evolutionary strategy-based optimisation algorithm (CMA-ES~\cite{hansen_cma_2023}) that attempts to find solutions that satisfy the requirements of safety whilst successfully navigating aircraft to their required exit conditions as efficiently as possible. Figure~\ref{fig:falcon} illustrates Falcon controlling simulated aircraft, with the optimised plans for each aircraft and waypoints at which clearances are planned, shown in magenta.

\section{Experimental Validation}

\subsection{Inter-Rater Reliability (IRR) Assessment}

\begin{figure}[t]
\centering
\begin{minipage}[t]{.45\textwidth}
  \centering
  \captionsetup{width=\linewidth}
  \includegraphics[width=.9\linewidth,trim={0pt 0pt 290pt 30pt}, clip]{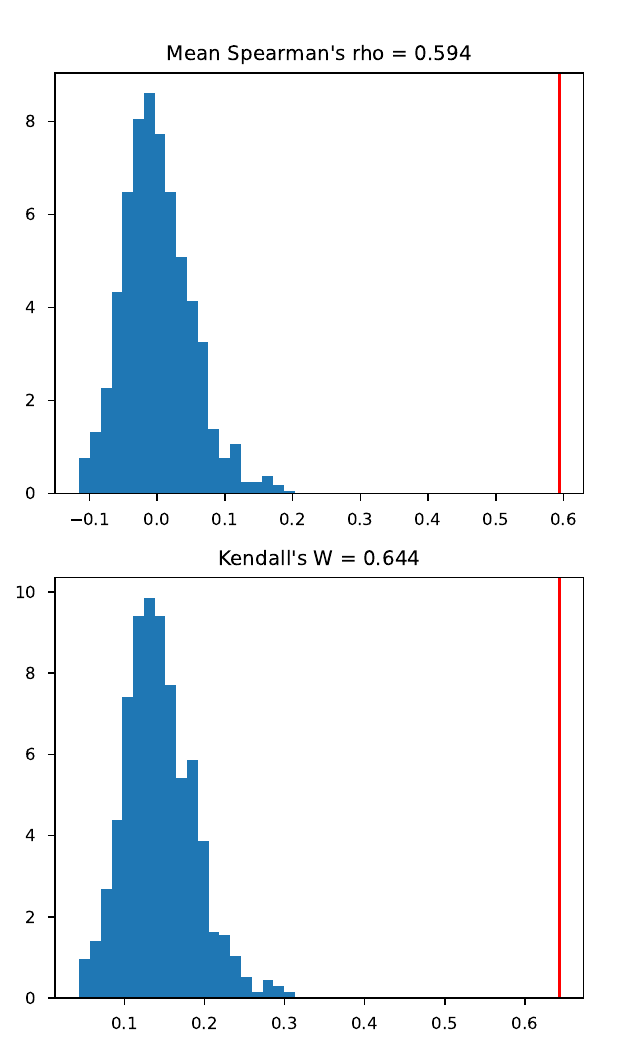}
  \captionof{figure}{Histograms of the distributions of  measures of inter-rater correlation for randomised scores.  The vertical red lines mark the unrandomised mean Spearman's rho (top) and Kendall's W (bottom) between instructor scores. }
  \label{fig:spearman}
\end{minipage}\hfill
\begin{minipage}[t]{.45\textwidth}
  \centering
  \captionsetup{width=\linewidth}
  \includegraphics[width=0.91\linewidth]{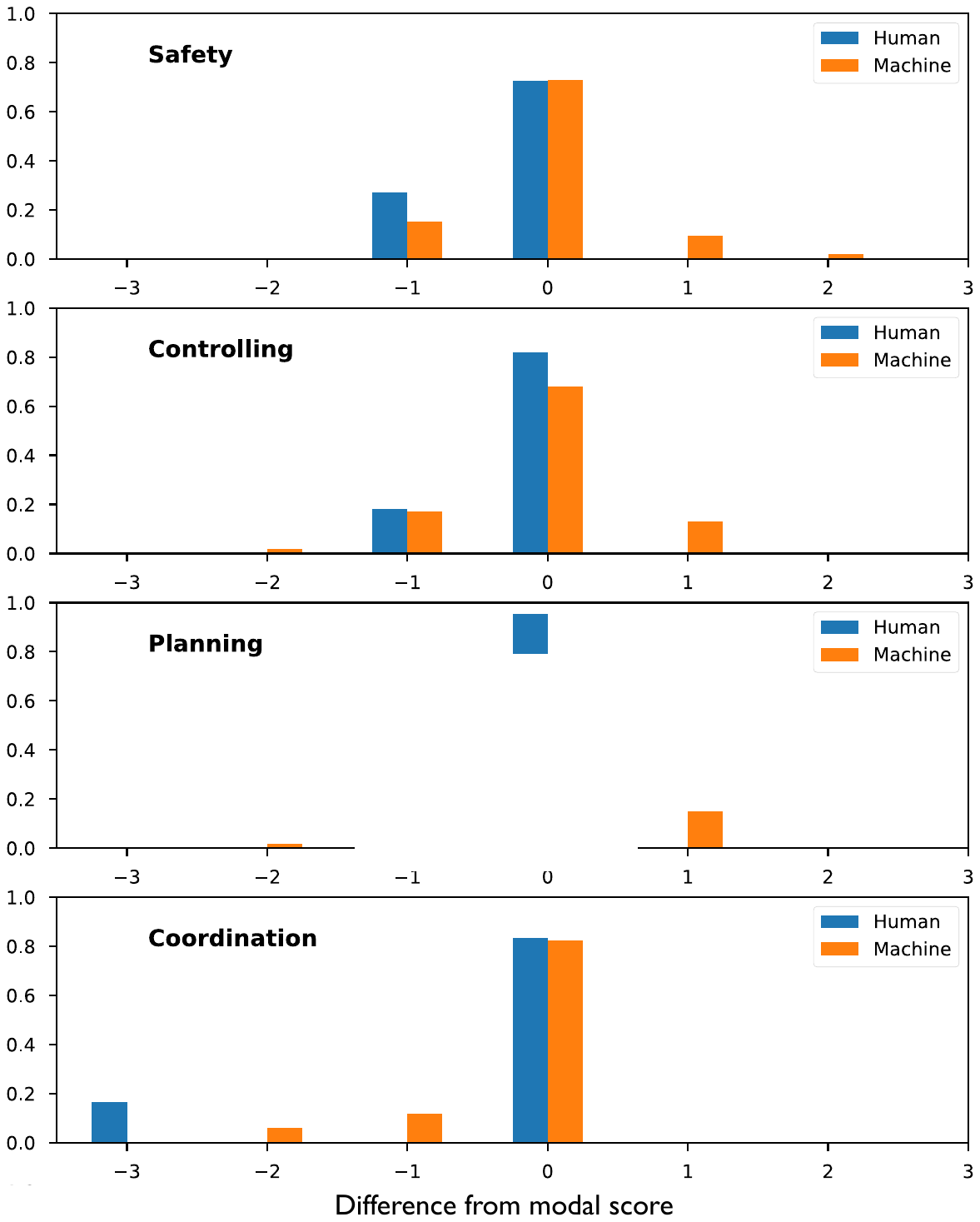}
  \captionof{figure}{Histograms showing the frequency of differences of individual instructors' scores from the consensus (modal) score.}
  \label{fig:deviations}

\end{minipage}
\end{figure}

To validate the suitability of the framework for assessing ATC competence, we performed an extended test for Inter-Rater Reliability (IRR). This measures how consistently ratings could be made within the BluebirdDT environment, utilising the MBT curriculum. A set of trainee data from the real-world college was obtained and converted to be replayed through the BluebirdDT environment, alongside replays of agent control. Inter-rater reliability was then assessed by inviting ATC college instructors to evaluate the same run and by comparing their scores. For this experiment, we mixed human trainee runs with agent runs, allowing for any issues with scoring agent behaviours to be identified. In total, we compared the scores on 19 scenarios (each with a duration of approximately 30 minutes) that were independently assessed by at least 7 instructors. IRR metrics were calculated using Intraclass Correlation (consistency and agreement), the mean Spearman's rho, and the rank-based measure of ordinal correlation, Kendall's W. As suggested by \citet{deRaadt21Reliability}, these all provided similar results. We report the mean Spearman's rho of 0.59 and Kendall's W of 0.64. To assess the significance of these figures, we randomly permuted the scores given by instructors for each scenario to remove the correlation between instructors and recalculated the correlation statistic; repeating the random permutation many times effectively destroys inter-instructor correlation while preserving the distribution of scores. Figure~\ref{fig:spearman} shows that the distributions of randomised scores are clearly distinct from the measured values of inter-rater reliability. Figure~\ref{fig:deviations} shows histograms of the differences between instructors' scores for each competency and the consensus (modal) score. It is clear that, for both human and machine agents, there is strong agreement among the assessors on how each performed. We particularly note that there is no appreciable difference in the consistency of scores for machine agents or human trainees.

\subsection{Simulator Fidelity and Realism}

The emulation of the formative and summative exercises in BluebirdDT is key to the Machine Basic framework. This is critical to ensure that the conflicts within the hand-crafted Basic exercises are not altered when they are simulated. The primary source of misspecification when emulating the college scenarios in BluebirdDT arises from discrepancies between BluebirdDT's trajectory predictor and that used by the simulators at NATS' training college.

To verify the accuracy of BluebirdDT's simulated trajectories, we use data logged from exercises performed by trainee ATCOs on NATS' training college simulators, including sequences of clearances issued and the characteristics of each scenario (e.g., wind direction, aircraft starting locations, and characteristics). This data is gathered from November 20th 2024 to June 20th 2025 for simulations that ran to completion. Aircraft that were in the background (i.e., those that do not enter the Medway sector) or that received clearances that BluebirdDT and the agents do not currently use were excluded as being out of scope. BluebirdDT takes as input the same clearance sequences and scenario characteristics, and the output trajectories are then compared against the college trajectory data. This process forms part of our complete digital twin assurance strategy, detailed in full by \citet{bluebird_assurance}.

\begin{figure}
    \centering
    \includegraphics[width=0.8\linewidth]{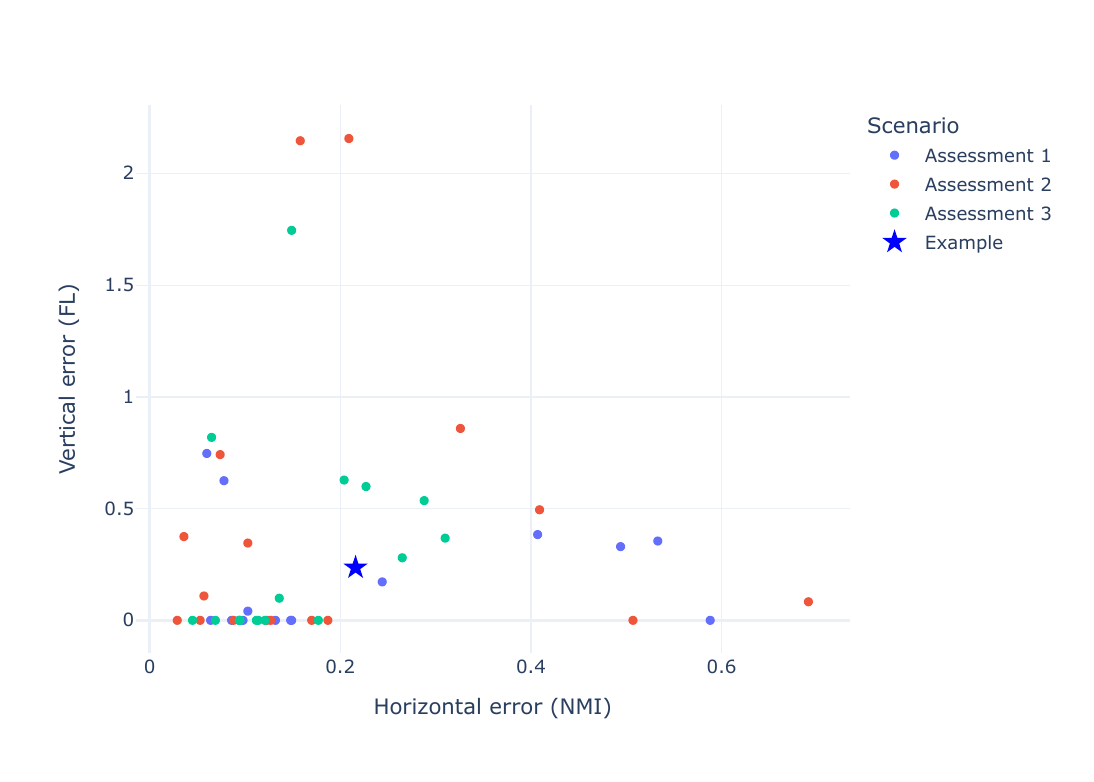}
    \caption{Average errors from replicating the 3 summative exercises used to assess the agents, coloured by the assessment scenario. Each point represents an aircraft, averaged over the simulations performed by the trainee ATCOs. The starred example is pictured in Figure \ref{fig:example_verification_aircraft}.}
    \label{fig:medway_verification}
\end{figure}

\begin{figure}
\centering
\begin{minipage}{0.4\linewidth}
   \centering
   \includegraphics[width=\linewidth]{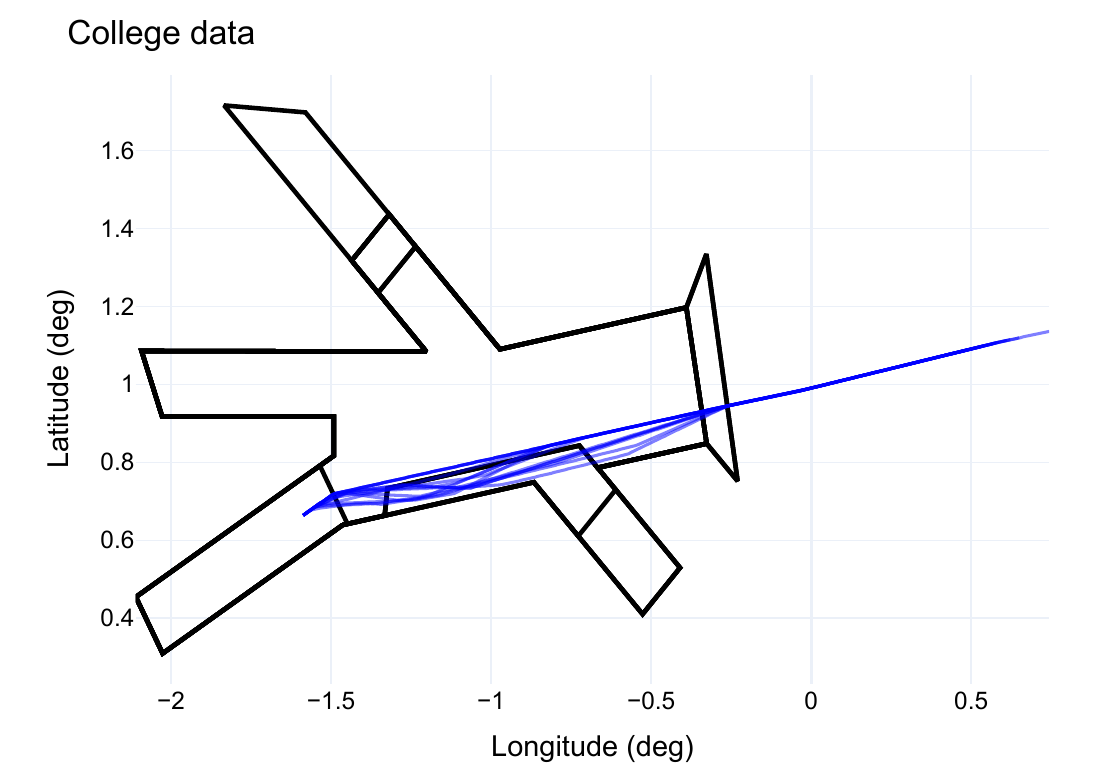}
\end{minipage}
\begin{minipage}{0.4\linewidth}
   \centering
   \includegraphics[width=\linewidth]{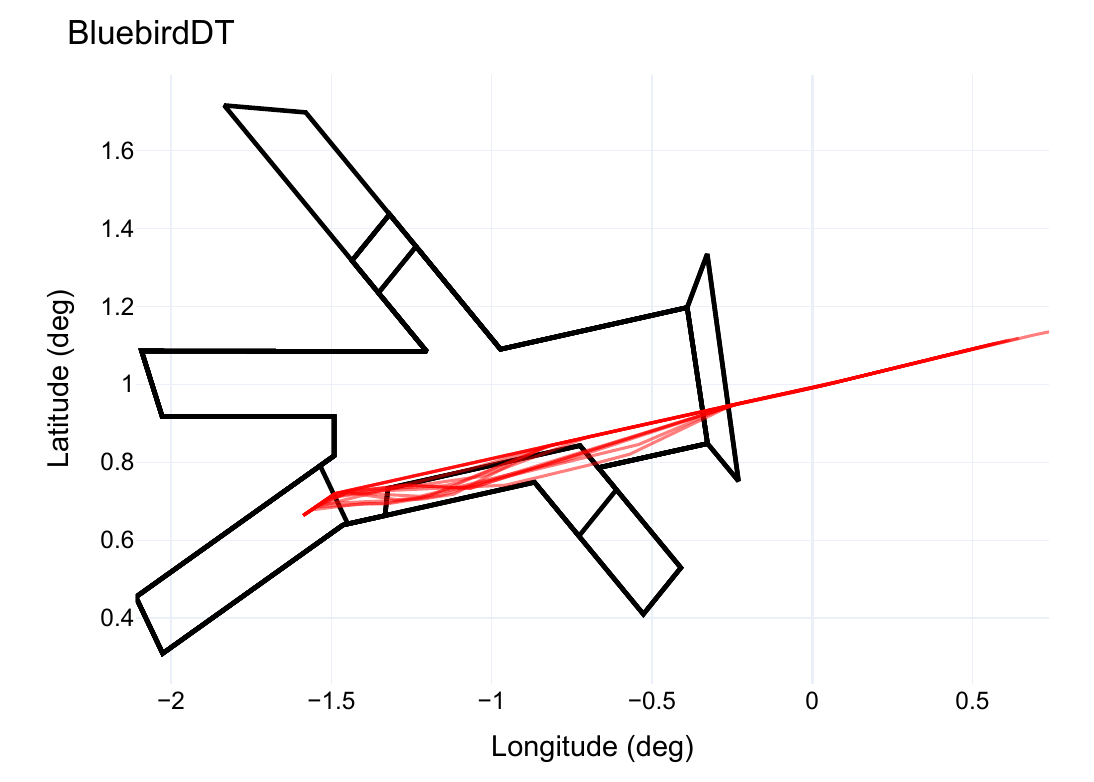}
\end{minipage}
\begin{minipage}{0.4\linewidth}
   \centering
   \includegraphics[width=\linewidth]{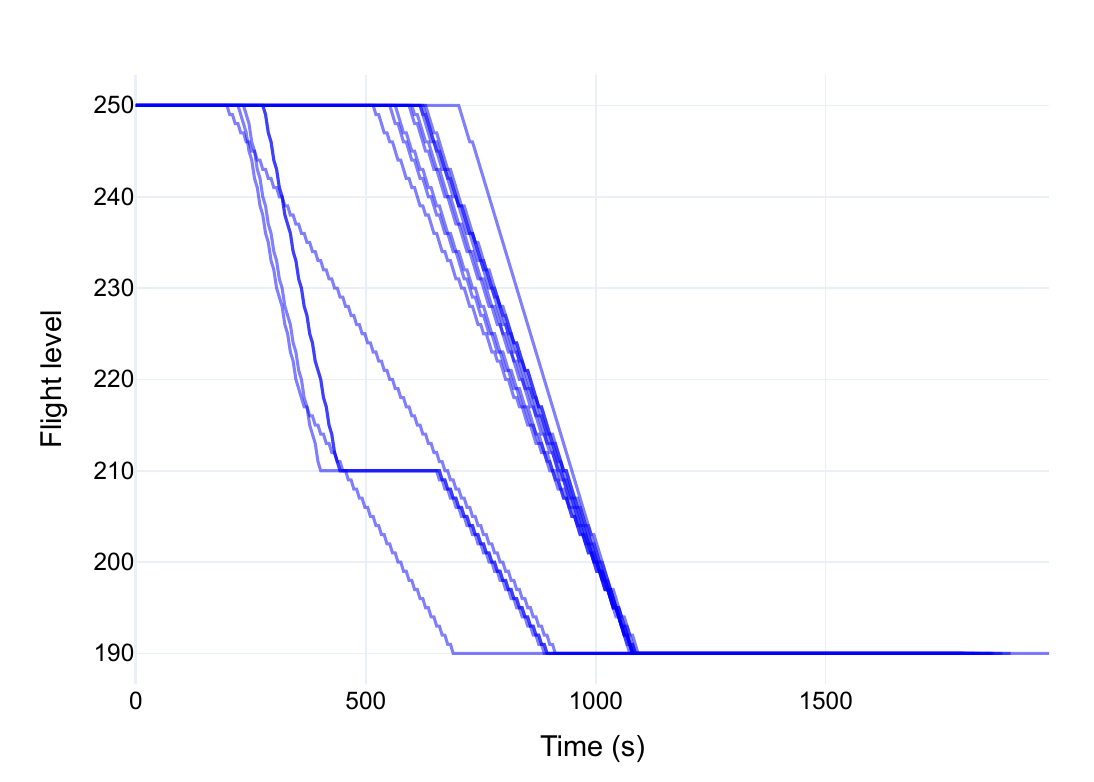}
\end{minipage}
\begin{minipage}{0.4\linewidth}
   \centering
   \includegraphics[width=\linewidth]{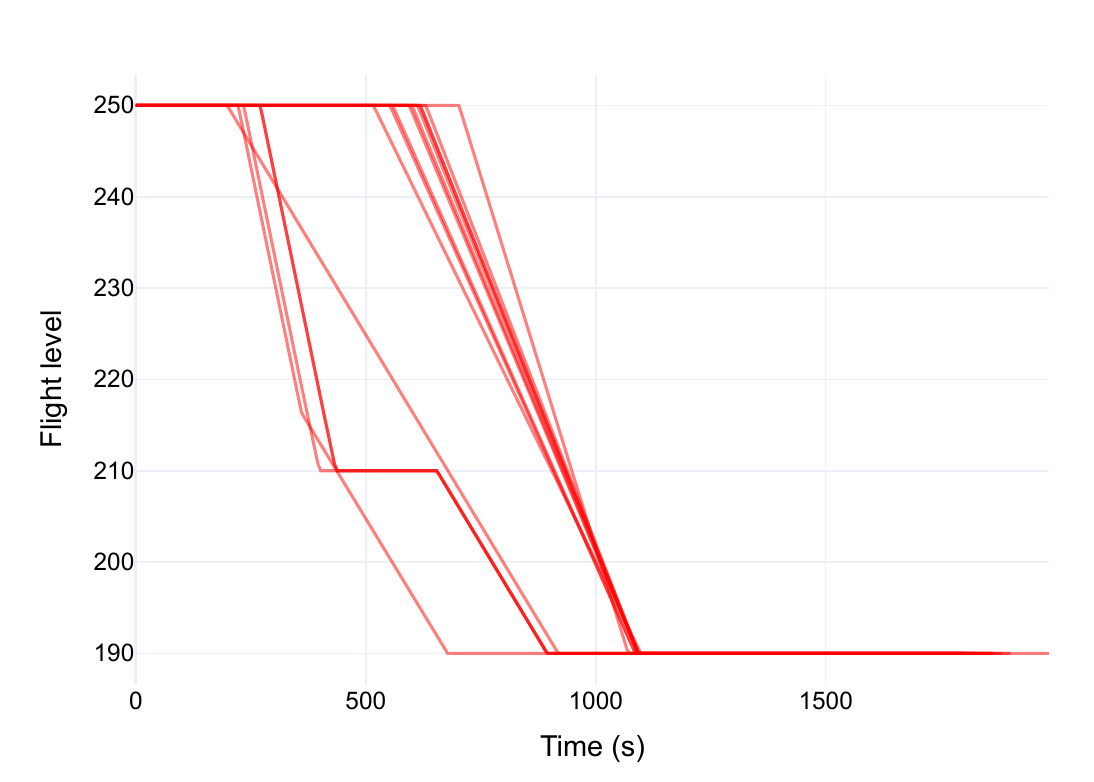}
\end{minipage}
\caption{Example of one aircraft in a summative scenario, over the course of 15 simulations. The left (blue) trajectories are those in the college data and the right (red) trajectories are those emulated by the trajectory predictor in BluebirdDT.}
\label{fig:example_verification_aircraft}
\end{figure}

Figure~\ref{fig:medway_verification} shows the horizontal and vertical errors for a given aircraft in a given scenario, averaged over all simulations. Figure~\ref{fig:example_verification_aircraft} zooms in on one of these aircraft (shown as a star in Figure~\ref{fig:medway_verification}): the left panels show the data (in blue), with each horizontal (top panel) and vertical trajectory (bottom panel) representing one training run for that aircraft. The right panels show the BluebirdDT simulations (in red), demonstrating an excellent emulation of this aircraft for varying sets of issued clearances. Table~\ref{table:medway_verification} shows the mean errors and their standard deviation (SD) over the simulations for a specific assessment. A maximum error of 5 flight levels vertically and 2.5 nautical miles laterally (i.e., half that of the separation standard) were used as threshold values for an acceptable simulation of an aircraft. These threshold values were suggested by instructors as discrepancies that were unlikely to alter the set of conflicts within the hand-crafted exercises. Table~\ref{table:medway_verification} shows the percentage of aircraft that exceeded this threshold, which is under $8\%$ for all assessments. Such aircraft were manually checked to confirm that their behaviour would not have a significant effect, such as causing an unintended conflict or removing an intended conflict.

\begin{table}[h]
    \centering
    \caption{Verification summary for each of the three assessments evaluated.}
    \begin{tabular}{ccccccc}
       \hline
       Assessment & Number of simulations & Aircraft in threshold (\%) & \multicolumn{2}{c}{Horizontal error (NM)} & \multicolumn{2}{c}{Vertical error (FL)} \\
        \cline{4-7}
          & & & Mean & SD & Mean & SD \\
        \hline
        1 & 18 & 100.0 & 0.25 & 0.04 & 0.10 & 0.04 \\
        2 & 16 & 97.9 & 0.26 & 0.03 & 0.32 & 0.22 \\
        3 & 12 & 92.1 & 0.17 & 0.02 & 0.37 & 0.11 \\
        \hline
    \end{tabular}
    \label{table:medway_verification}
\end{table}

\section{Results and Discussion}

\subsection{Agent Assessment Agreement}

\begin{figure} [!h]
    \centering
    \includegraphics[width=1.0\textwidth]{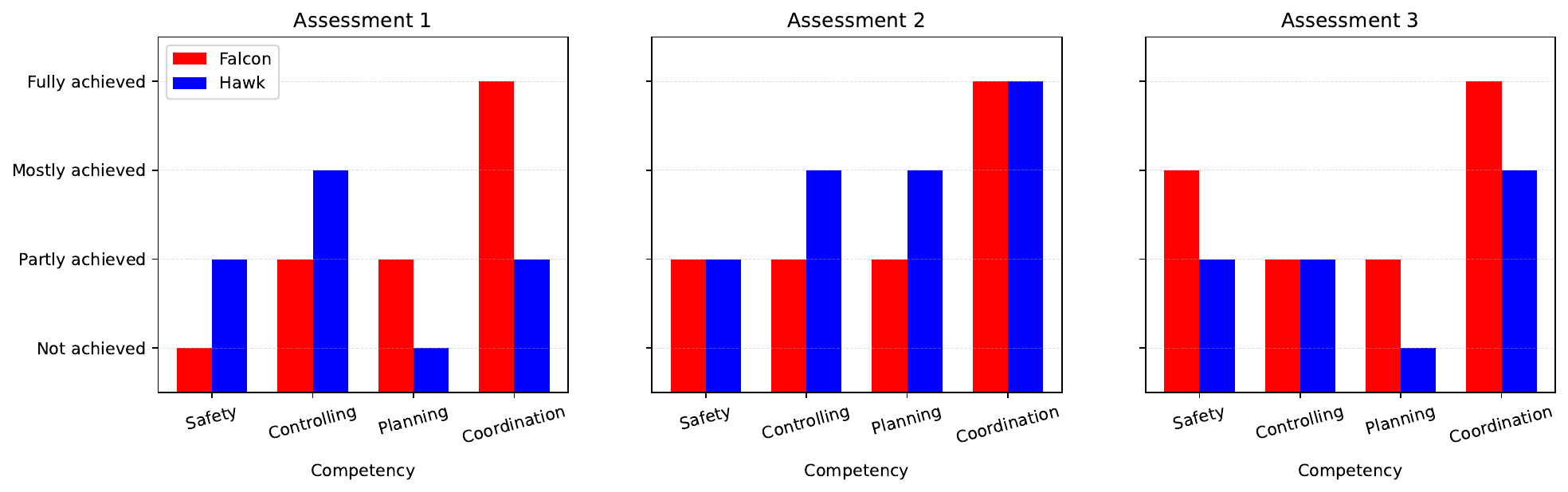}
    \caption{Summative results round 1 for Falcon and Hawk.}
    \label{fig:results_round_1} 
\end{figure}

Both candidate agents were successfully tested against the assessment framework. An initial round of testing was conducted in January 2025 to set baseline performance. A total of three 30-minute assessment runs were executed for each agent, with the results considered and moderated by a separate assessor in line with college procedure. The results from the individual runs in the first trial can be seen in Figure~\ref{fig:results_round_1}. Both agents were able to score more than the minimum in all areas assessed, showing the beginnings of traction on the assessed task. However, the overall assessor gradings in Table~\ref{tab:assessor_grades} were mostly unsatisfactory, indicating that the agents had not yet met the expected standard for the MBT curriculum. Real students are expected to achieve mostly full marks in each exercise, with any competencies marked below this being carefully considered when making the final assessment. Due to its expert-informed origins, the rules-based agent Hawk scores more highly in the controlling category, as the solutions it deploys have been, in part, formed through expert elicitation, leading to more naturally recognisable control techniques. The optimisation-based agent Falcon produces solutions that are less aligned with expected taught techniques, as it is more likely to seek highly optimised paths over ordered and repeatable behaviours. However, as achieving the required sector exit coordination is set as a constraint in the problem formulation, Falcon scores perfectly in this category. Safety for both agents requires further work, but it should be noted that a single perceived safety error is often enough to severely reduce grading on a summative assessment. Safety is paramount in ATC, and this is reflected even in these early stages of training, with students (and, by extension, agents) expected to demonstrate astute adherence to complex and nuanced definitions of safety.

A key part of the assessment process is the generation of grading forms, which detail the rationale for the scores given. For instance, in the assessor summary comments for Hawk, the performance in Coordination was summarised as:

\begin{quoting}
    \textit{Three examples across two summative assessments in which aircraft did not achieve their exit levels. One level was missed significantly involving two other sectors above. No attempt to orbit this aircraft and get the height off was made.}
\end{quoting}

This form of feedback, along with more detailed per-run write-ups, allows agent developers to understand the weaknesses and areas for development in their approaches.

\begin{figure} [!h]
    \centering
    \includegraphics[width=1.0\textwidth]{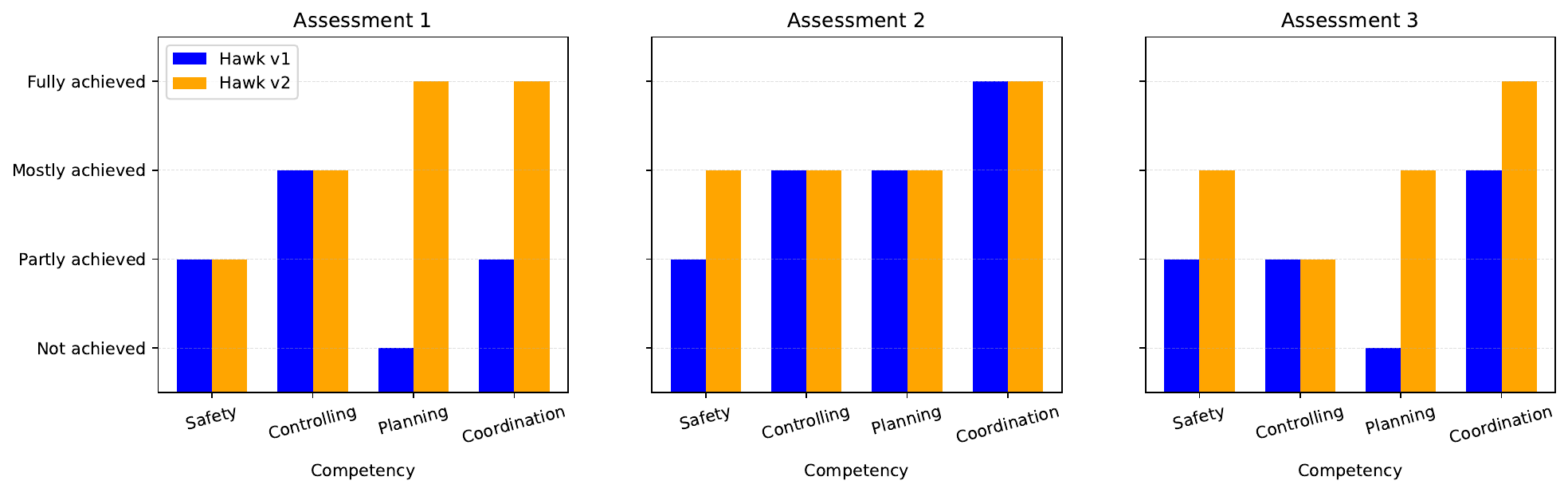}
    \caption{Summative results for Hawk v1 vs Hawk v2.}
    \label{fig:old_hawk_new_hawk} 
\end{figure}

Following revisions to the implemented logic, agent Hawk was re-tested in August 2025, with the individual run progress shown in Figure~\ref{fig:old_hawk_new_hawk}. The final assessor grading in Table~\ref{tab:assessor_grades} reflects its increased consistency and performance across all measured competencies, where Hawk now scores a \textit{Satisfactory} grade across all competency areas except for Safety. Following targeted improvements to prioritising efficient achievement of exit levels, the summary comments for Hawk on Coordination for this second round were:

\begin{quoting}
    \textit{Across all 3 summative exercises, all aircraft achieved their coordinated exit levels and no re-coordination was required.
Therefore, this objective is graded as satisfactory.}
\end{quoting}

Similar improvements were observed in planning. The summary comments from round 1 noted that:

\begin{quoting}
    \textit{...it appeared that conflicting traffic was known with short term plans put in place. It was not always evident that the bigger picture was taken into account and plans did not always feel fully developed.}
\end{quoting}

Following this feedback, the internal logic of Hawk was modified to identify conflicts earlier and pre-position aircraft to allow for more effective control. These efforts were then reflected in the summary comments for planning from round 2:

\begin{quoting}
    \textit{...it was noted that forward planning was taking place with conflicts detected so that a plan of action to resolve them could be made. This was evidenced by aircraft being positioned early to allow for climbs and descents to resolve conflicts or safe initial levels used until it was confirmed that further safe levels could be used. Therefore, a consistent method of operation has been demonstrated to grade this objective as satisfactory.}
\end{quoting}

The most important area for development remains Safety, and the feedback obtained for both agents highlights the importance of the MBT in elucidating domain-driven measures of success. The summary comments for Falcon noted:

\begin{quoting}
    \textit{There were no losses of separation in any of the exercises however there were multiple scenarios where safety was not ensured.}
\end{quoting}

This echoes comments for Hawk in the same area:

\begin{quoting}
    \textit{Whilst separation was maintained throughout, one example of an unsafe clearance (with retrospective headings applied) and six examples of failures to ensure separation were noted.}
\end{quoting}

The standard for safety, even at this early training stage, is founded on the notion of \textit{ensuring separation}. This is a process whereby the issued clearances are fail-safe; it goes beyond predicting future conflicts purely from current trajectories and embeds an appreciation of operational uncertainty to produce plans that are demonstrably safe, regardless of variations in weather, pilot response time, or any other relevant source of uncertainty.

By including experts in our assessment process, the feedback gathered provides both positive and negative feedback loops. The detail provided by expert assessors in summative reports allows for the identification of desirable behaviours for reinforcement, alongside a clear description of issues to be solved. Detail beyond coarse numerical measures allows agent designers to take steps to codify measures of ATC performance that have not previously been available, either from industry or academic research.

As our work continues, we will develop numerical objectives and reward functions, refined and tested against expert feedback, with the emergent behaviours of agents designed with these reward functions serving both to validate their performance and to allow for expert-in-the-loop modification towards a true representation of ATC success (initial work on this is presented in \citep{bluebird_agent_transparency}). In this way, we begin to close the gap between academic representations of ATC and the true domain task.

\FloatBarrier

\FloatBarrier

\begin{table}[ht]
\centering
\caption{Overall Assessor Gradings - Assessment Round 1 to Round 2}
\renewcommand{\arraystretch}{1.2}
\begin{tabular}{l c c c}
\toprule
\textbf{Competency} & \textbf{Falcon Round 1 (Jan 2025)} & \textbf{Hawk Round 1 (Jan 2025)} & \textbf{Hawk Round 2 (Aug 2025)} \\
\midrule
Safety        & Unsatisfactory & Unsatisfactory & Unsatisfactory \\
Controlling   & Unsatisfactory & Unsatisfactory & \textbf{Satisfactory} $\uparrow$ \\
Planning      & Unsatisfactory & Unsatisfactory & \textbf{Satisfactory} $\uparrow$ \\
Coordination  & \textbf{Satisfactory} & Unsatisfactory & \textbf{Satisfactory} $\uparrow$ \\
\bottomrule
\end{tabular}
\label{tab:assessor_grades}
\end{table}

\section{Future Work}\label{sec:future_work}

Future work on this framework will focus on improving the fidelity of assessments possible through refinements to the BluebirdDT platform, adding functional support for a range of procedures, services outside controlled airspace, airport departure management, and other key ATC concepts that pose challenges for ATC both in later training and in real-world operations. These new functional areas will necessitate further work in mapping the required competencies and objectives for agents, to facilitate an effective design process. We will engage the regulator to advance conversations on the potential for establishing more formal frameworks for assurance, and embed the knowledge and expertise of instructors and operational staff in effective numerical measures of success. 

As part of the work of Project Bluebird, we will soon release an open-source airspace sector of similar fidelity to the proprietary training airspace we have used so far. This open-source release will form the foundation of an open curriculum, providing a range of pre-designed and automatically generated test scenarios. This shift will also allow for the design of scenarios built for use with probabilistic aircraft modelling\cite{bluebird_data_assim} to incorporate real-world behaviours, as this is already supported within BluebirdDT. We are now strongly positioned to iterate on the design of numerical measures, as the emergent behaviour of agents incentivised by these measures can be directly demonstrated and assessed by experts. Through this mechanism of action, we hope to soon provide an open, comprehensive training and assessment solution for a broad family of agent approaches, effectively aligning the true industrial problem with the academic vanguard.

\section{Conclusion}

We have demonstrated a new assessment framework for automated approaches in ATC. By utilising the legally justified and regulated college curriculum, we directly align our standards  with the fundamentals of the task and help to set a more challenging and realistic direction for agent design. Trust in future automated solutions will be reliant on alignment with the true controlling task, and this framework provides a mechanism for iterating towards this goal.

\section{Appendix: MBT Objectives Scope}\label{sec:mbt_appendix}

\input{MBT_curriculum}

\section*{Acknowledgments}
This work was supported by the EPSRC (EP/V056522/1) and NATS.

\bibliography{references_all}

\end{document}

%% file: MBT_curriculum.tex
\begin{longtable}{|p{1.2cm}|p{3.0cm}|p{2.2cm}|p{4.8cm}|p{3.0cm}|}
    
    \caption{Air Traffic Control Competency and Scope Breakdown for MBT. The performance criteria are primarily drawn from the ICAO Manual on Air Traffic Controller Competency-Based Training and Assessment \cite{icao10056}}. \label{tab:competency_scope} \\
    
    \hline 
    \textbf{Scope} & \textbf{Identifier} & \textbf{Competency} & \textbf{Performance Criteria} & \textbf{Notes} \\
    \hline
    \endfirsthead

    \multicolumn{5}{c}%
    {\tablename\ \thetable\ -- \textit{Continued from previous page}} \\
    \hline
    \textbf{Scope} & \textbf{Identifier} & \textbf{Competency} & \textbf{Performance Criteria/Observable Behaviour} & \textbf{Notes} \\
    \hline
    \endhead

    \hline
    \multicolumn{5}{r}{\textit{Continued on next page}} \\
    \endfoot

    \hline
    \endlastfoot
    
    In Scope & MBT.SAFETY.001 & Safety & Ensures that safe separation between aircraft is always maintained & Primary area of focus for agent development \\
    In Scope & MBT.SAFETY.002 & Safety & Reacts appropriately to situations that have the potential to become unsafe & \\
    In Scope & MBT.SAFETY.003 & Safety & Understands and applies appropriate techniques to ensure lateral separation when required & \\
    In Scope & MBT.SAFETY.004 & Safety & Understands and applies appropriate techniques to ensure vertical separation when required & \\
    In Scope & MBT.SAFETY.005 & Safety & Implements appropriate solutions to catch up conflict scenarios & \\
    In Scope & MBT.SAFETY.006 & Safety & Implements appropriate solutions to crossing track conflict scenarios & \\
    In Scope & MBT.SAFETY.007 & Safety & Implements appropriate solutions to reciprocal conflict scenarios & \\
    In Scope & MBT.PLAN.001 & Planning & Analyses and interprets the situation based on information acquired from monitoring and scanning available data sources and tools & \\
    In Scope & MBT.PLAN.002 & Planning & Predicts the future operational situation & \\
    In Scope & MBT.PLAN.003 & Planning & Identifies potentially hazardous situations (e.g. amount of separation with other aircraft, navigational deviations) & \\
    In Scope & MBT.PLAN.004 & Planning & Identifies traffic conflicts (aircraft that will interact in the future) & \\
    In Scope & MBT.PLAN.005 & Planning & Takes into account the existing rules and operating procedures when determining possible solutions to a problem & \\
    In Scope & MBT.PLAN.006 & Planning & Plans safe conflict resolutions & \\
    In Scope & MBT.PLAN.007 & Planning & Implements an appropriate solution to a problem & \\
    In Scope & MBT.PLAN.008 & Planning & Applies appropriate air traffic separation and spacing & \\
    In Scope & MBT.PLAN.009 & Planning & Organizes tasks in accordance with an appropriate order of priorities & \\
    In Scope & MBT.PLAN.010 & Planning & Issues clearances in a timely manner & \\
    In Scope & MBT.COORD.001 & Coordination & Ensures that when aircraft leave the sector airspace it is in accordance with the terms of the agreed coordination & \\
    In Scope & MBT.COORD.002 & Coordination & Revises coordinated agreements in a timely manner when the required exit conditions cannot be achieved & All levels are intended to be achievable on Basic \\
    In Scope & MBT.CONTROL.001 & Controlling & Monitors air traffic in own area of responsibility and nearby airspace & \\
    In Scope & MBT.CONTROL.002 & Controlling & Acquires information from available surveillance a flight data systems, meteorological data, electronic data displays and any other means available & \\
    In Scope & MBT.CONTROL.003 & Controlling & Integrates information acquired from monitoring and scanning into the overall picture & \\
    In Scope & MBT.CONTROL.004 & Controlling & Understands the airspace structure and its implications for effective and safe control, such as distance-inside requirements below FL195 & \\
    In Scope & MBT.CONTROL.005 & Controlling & Manages arriving, departing, and/or en-route traffic using prescribed procedures & \\
    In Scope & MBT.CONTROL.006 & Controlling & Takes responsibility for own performance, detecting and resolving own errors & \\
    In Scope & MBT.CONTROL.007 & Controlling & Ensures that traffic on vectors is suitably positioned for the next sector, taking into account onward routings and destinations & \\
    In Scope & MBT.CONTROL.008 & Controlling & Understand deemed route separation where applicable & \\
    In Scope & MBT.CONTROL.009 & Controlling & Takes aircraft performance into account when issuing clearances & \\
    In Scope & MBT.CONTROL.010 & Controlling & Selects the most appropriate separation method & \\
    In Scope & MBT.CONTROL.011 & Controlling & Issues clearances that take into account aircraft performance and airspace constraints & \\
    In Scope & MBT.CONTROL.012 & Controlling & Issues clearances that ensure separation is maintained & \\
    In Scope & MBT.CONTROL.013 & Controlling & Issues clearances that resolve traffic conflicts & \\
    In Scope & MBT.CONTROL.014 & Controlling & Monitors the execution of separation actions & \\
    In Scope & MBT.CONTROL.015 & Controlling & Adjusts control actions, when necessary, to maintain separation & \\
    In Scope & MBT.CONTROL.016 & Controlling & Considers timeliness in decision-making & \\
    In Scope & MBT.CONTROL.017 & Controlling & Uses a variety of techniques to effectively manage the traffic (e.g. vectoring, trade-off levels) & \\
    In Scope & MBT.CONTROL.018 & Controlling & Maintains focus despite varying traffic levels & \\
    In Scope & MBT.CONTROL.019 & Controlling & Issues clearances to flight crew that result in an efficient traffic flow & \\
    In Scope & MBT.COMMS.001 & Communication & Transfers communication to the next sector in a timely manner & \\
    Partial & MBT.CONTROL.020 & Controlling & Accounts for wind when executing clearances & Wind is configurable in the digital twin, and can be disabled if necessary dependent on agent capabilities \\
    Partial & MBT.COORD.003 & Coordination & Coordinates with personnel in other operational positions and other stakeholders in a timely manner & Coordination is fixed and pre-populated \\
    Partial & MBT.COORD.004 & Coordination & Coordinates the movement, control and transfer of control for flights using the prescribed coordination procedures & Coordination is fixed and pre-populated \\
    Out of Scope & MBT.SAFETY.008 & Safety & Issues traffic proximity information to flight crews in a relevant, accurate and timely manner & \\
    Out of Scope & MBT.PLAN.011 & Planning & Maintains an up-to-date flight progress strip display, ensuring all relevant control and coordination information is recorded and updated & A rudimentary strip board is provided in the digital twin, and is automatically updated \\
    Out of Scope & MBT.PLAN.012 & Planning & Represents conflicts in the flight progress strip display & Strip bay is automatically updated without agent involvement \\
    Out of Scope & MBT.COORD.005 & Coordination & Coordinates with the next sector when it is necessary to transfer aircraft on headings & Coordination is to be assumed for headings \\
    Out of Scope & MBT.COORD.006 & Coordination & Provides a radar handover where necessary to other agencies & \\
    Out of Scope & MBT.COORD.007 & Coordination & Resolves conflicts through coordination with adjacent sectors or units & \\
    Out of Scope & MBT.COORD.008 & Coordination & Uses clear and concise terminology for verbal coordination & \\
    Out of Scope & MBT.CONTROL.021 & Controlling & Verifies the identity of aircraft arriving from other agencies through the use of squawk change, squawk ident, position report, or turn method as appropriate & \\
    Out of Scope & MBT.CONTROL.022 & Controlling & Changes the service of aircraft which leave controlled airspace & \\
    Out of Scope & MBT.CONTROL.023 & Controlling & Understands and uses correct directional levels for aircraft travelling East or West & \\
    Out of Scope & MBT.CONTROL.024 & Controlling & Understands and applies the criteria for level vacation, to expedite the traffic flow & \\
    Out of Scope & MBT.CONTROL.025 & Controlling & Verifies the mode C information of any aircraft arriving from another agency & \\
    Out of Scope & MBT.CONTROL.026 & Controlling & Applies control methods to manage non-transponding aircraft & Controlled primary-only aircraft are represented as transponder equipped in the digital twin. \\
    Out of Scope & MBT.COMMS.002 & Communication & Uses standard radiotelephony phraseology, when prescribed & Radiotelephony is emulated using text only, with a fixed action set and syntax \\
    Out of Scope & MBT.COMMS.003 & Communication & Prioritises transmissions to account for serial transmission of instructions & The digital twin permits simultaneous transmission \\
    Out of Scope & MBT.COMMS.004 & Communication & Transfers aircraft to the appropriate onward frequency & \\
    Out of Scope & MBT.COMMS.005 & Communication & Provides necessary routing information to aircraft entering the sector from other agencies & \\
    Out of Scope & MBT.COMMS.006 & Communication & Speaks clearly, accurately and concisely & \\
    Out of Scope & MBT.COMMS.007 & Communication & Uses appropriate vocabulary and expressions to convey clear messages & \\
    Out of Scope & MBT.COMMS.008 & Communication & Adjusts speech techniques to suit the situation & \\
    Out of Scope & MBT.COMMS.009 & Communication & Verifies accuracy of read backs and corrects as necessary & \\
    Out of Scope & MBT.COMMS.010 & Communication & Instructs aircraft on headings to report the heading when transferring communications to the next sector & \\
\end{longtable}

%% file: main.bbl
\begin{thebibliography}{29}
\newcommand{\enquote}[1]{``#1''}
\providecommand{\natexlab}[1]{#1}
\providecommand{\url}[1]{\texttt{#1}}
\providecommand{\urlprefix}{URL }
\expandafter\ifx\csname urlstyle\endcsname\relax
  \providecommand{\doi}[1]{\discretionary{}{}{}https://doi.org/#1}\else
  \providecommand{\doi}[1]{\discretionary{}{}{}\urlstyle{rm}\url{https://doi.org/#1}}\fi

\bibitem[{Lymperopoulos and Lygeros(2010)}]{lymperopoulos2010sequential}
Lymperopoulos, I., and Lygeros, J., \enquote{{Sequential Monte Carlo methods
  for multi-aircraft trajectory prediction in air traffic management},}
  \emph{International Journal of Adaptive Control and Signal Processing},
  Vol.~24, No.~10, 2010, pp. 830--849.
\newblock \urlprefix\url{{https://doi.org/10.1002/acs.1174}}.

\bibitem[{{CAA}(2024)}]{CAP493}
{CAA}, \emph{{CAP} 493: {Manual} of {Air} {Traffic} {Services} ({MATS}) {Part}
  1 - {Civil} {Aviation} {Authority}}, 2024.
\newblock
  \urlprefix\url{https://www.caa.co.uk/our-work/publications/documents/content/cap-493/},
  {A}ccessed: 26-06-2024.

\bibitem[{Wesson(1977)}]{wesson_planning_1977}
Wesson, R.~B., \enquote{Planning in the world of the air traffic controller,}
  \emph{Proceedings of the 5th international joint conference on {Artificial}
  intelligence - {Volume} 1}, Morgan Kaufmann Publishers Inc., San Francisco,
  CA, USA, 1977, pp. 473--479.

\bibitem[{{J-P Nicolaon, G. Dean} and Miller(1995)}]{ARC2000}
{J-P Nicolaon, G. Dean}, and Miller, W., \enquote{An overview of {ARC} 2000
  {Version} 3 from the operational point of view,} EUROCONTROL, 1995.
\newblock \urlprefix\url{https://www.eurocontrol.int/node/10207}, {A}ccessed:
  26-06-2024.

\bibitem[{Erzberger(2006)}]{erzberger_automated_2006}
Erzberger, H., \enquote{Automated conflict resolution for air traffic control,}
  \emph{International Congress of the Aeronautical Sciences}, 2006.

\bibitem[{Brittain and Wei(2019)}]{brittain_autonomous_2019}
Brittain, M., and Wei, P., \enquote{Autonomous {Air} {Traffic} {Controller}:
  {A} {Deep} {Multi}-{Agent} {Reinforcement} {Learning} {Approach},}
  \emph{International {Conference} on {Machine} {Learning}: {ICML} 2019 {RL}
  for {Real} {Life} {Workshop}}, 2019.
\newblock \urlprefix\url{http://arxiv.org/abs/1905.01303}.

\bibitem[{Sui et~al.(2023)Sui, Ma, and Dong}]{sui_conflict_2023}
Sui, D., Ma, C., and Dong, J., \enquote{Conflict {Resolution} {Strategy}
  {Based} on {Deep} {Reinforcement} {Learning} for {Air} {Traffic}
  {Management},} \emph{Aviation}, Vol.~27, No.~3, 2023, pp. 177--186.
\newblock \doi{10.3846/aviation.2023.19720},
  \urlprefix\url{https://journals.vilniustech.lt/index.php/Aviation/article/view/19720}.

\bibitem[{Wang et~al.(2019)Wang, Li, Wang, and Shen}]{wang_deep_2019}
Wang, Z., Li, H., Wang, J., and Shen, F., \enquote{Deep reinforcement learning
  based conflict detection and resolution in air traffic control,} \emph{IET
  Intelligent Transport Systems}, Vol.~13, No.~6, 2019, pp. 1041--1047.
\newblock \doi{10.1049/iet-its.2018.5357},
  \urlprefix\url{https://onlinelibrary.wiley.com/doi/10.1049/iet-its.2018.5357}.

\bibitem[{Pappas et~al.(1996)Pappas, Tomlin, and Sastry}]{pappas_conflict_1996}
Pappas, G., Tomlin, C., and Sastry, S., \enquote{Conflict resolution for
  multi-agent hybrid systems,} \emph{Proceedings of 35th {IEEE} {Conference} on
  {Decision} and {Control}}, Vol.~2, 1996, pp. 1184--1189 vol.2.
\newblock \doi{10.1109/CDC.1996.572644},
  \urlprefix\url{https://ieeexplore.ieee.org/document/572644}, iSSN: 0191-2216.

\bibitem[{Tumer and Agogino(2009)}]{tumer_improving_2009}
Tumer, K., and Agogino, A., \enquote{Improving {Air} {Traffic} {Management}
  with a {Learning} {Multiagent} {System},} \emph{IEEE Intelligent Systems},
  Vol.~24, No.~1, 2009, pp. 18--21.
\newblock \doi{10.1109/MIS.2009.10},
  \urlprefix\url{http://ieeexplore.ieee.org/document/4763650/}.

\bibitem[{EUROCONTROL(2023)}]{argos}
EUROCONTROL, \emph{{ARGOS} {Factsheet}}, Mar. 2023.
\newblock
  \urlprefix\url{https://www.eurocontrol.int/publication/argos-factsheet},
  {A}ccessed: 26-06-2024.

\bibitem[{Crootof et~al.(2022)Crootof, Kaminski, and
  Price~II}]{crootof_humans_2022}
Crootof, R., Kaminski, M.~E., and Price~II, W.~N., \enquote{Humans in the
  {Loop},} {Social Science Research Network}, Mar. 2022.
\newblock \doi{10.2139/ssrn.4066781},
  \urlprefix\url{https://papers.ssrn.com/abstract=4066781}.

\bibitem[{Wu et~al.(2022)Wu, Xiao, Sun, Zhang, Ma, and He}]{wu_survey_2022}
Wu, X., Xiao, L., Sun, Y., Zhang, J., Ma, T., and He, L., \enquote{A survey of
  human-in-the-loop for machine learning,} \emph{Future Generation Computer
  Systems}, Vol. 135, 2022, pp. 364--381.
\newblock \doi{10.1016/j.future.2022.05.014},
  \urlprefix\url{https://www.sciencedirect.com/science/article/pii/S0167739X22001790}.

\bibitem[{Kung et~al.(2023)Kung, Cheatham, Medenilla, Sillos, De~Leon,
  Elepaño, Madriaga, Aggabao, Diaz-Candido, Maningo, and
  Tseng}]{kung_performance_2023}
Kung, T.~H., Cheatham, M., Medenilla, A., Sillos, C., De~Leon, L., Elepaño,
  C., Madriaga, M., Aggabao, R., Diaz-Candido, G., Maningo, J., and Tseng, V.,
  \enquote{Performance of {ChatGPT} on {USMLE}: {Potential} for {AI}-assisted
  medical education using large language models,} \emph{PLOS Digital Health},
  Vol.~2, No.~2, 2023, p. e0000198.
\newblock \doi{10.1371/journal.pdig.0000198},
  \urlprefix\url{https://pmc.ncbi.nlm.nih.gov/articles/PMC9931230/}.

\bibitem[{Katz et~al.(2024)Katz, Bommarito, Gao, and
  Arredondo}]{katz_gpt-4_nodate}
Katz, D.~M., Bommarito, M.~J., Gao, S., and Arredondo, P., \enquote{{GPT}-4
  passes the bar exam,} \emph{Philosophical transactions. Series A,
  Mathematical, physical, and engineering sciences}, Vol. 382, No. 2270, 2024,
  p. 20230254.
\newblock \doi{10.1098/rsta.2023.0254},
  \urlprefix\url{https://pmc.ncbi.nlm.nih.gov/articles/PMC10894685/}.

\bibitem[{Riedmaier et~al.(2020)Riedmaier, Ponn, Ludwig, Schick, and
  Diermeyer}]{riedmaier_survey_2020}
Riedmaier, S., Ponn, T., Ludwig, D., Schick, B., and Diermeyer, F.,
  \enquote{Survey on {Scenario}-{Based} {Safety} {Assessment} of {Automated}
  {Vehicles},} \emph{IEEE Access}, Vol.~8, 2020, pp. 87456--87477.
\newblock \doi{10.1109/ACCESS.2020.2993730},
  \urlprefix\url{https://ieeexplore.ieee.org/document/9090897/}.

\bibitem[{Pepper et~al.(2026{\natexlab{a}})Pepper, Keane, Hodgkin, Dewi~Gould,
  Lauritsen, Vlahos, {De Ath}, Everson, Cannon, Castro, Korna, Carvell, and
  Thomas}]{bluebird_dt}
Pepper, N., Keane, A., Hodgkin, A., Dewi~Gould, E.~H., Lauritsen, L., Vlahos,
  C., {De Ath}, G., Everson, R., Cannon, R., Castro, A.~S., Korna, J., Carvell,
  B., and Thomas, M., \emph{A Probabilistic Digital Twin of UK En Route
  Airspace for Training and Evaluating AI Agents for Air Traffic Control},
  AIAA, 2026{\natexlab{a}}.

\bibitem[{{NATS}(2013)}]{NATS_ATCO_pathway}
{NATS}, \emph{Becoming a NATS controller}, 2013.
\newblock
  \urlprefix\url{{https://nats.aero/blog/wp-content/uploads/2013/06/Infographic_BecomingAController.pdf}}.

\bibitem[{{CAA}(2021{\natexlab{a}})}]{CAP413}
{CAA}, \emph{{CAP} 413: {Radiotelephony} {Manual} - {Civil} {Aviation}
  {Authority}}, 2021{\natexlab{a}}.
\newblock
  \urlprefix\url{https://www.caa.co.uk/our-work/publications/documents/content/cap-413/},
  {A}ccessed: 26-06-2024.

\bibitem[{{CAA}(2021{\natexlab{b}})}]{CAAreg2015}
{CAA}, \emph{{ATCO} {Air} {Traffic} {Controllers} {\textbar} {UK} {Civil}
  {Aviation} {Authority}}, 2021{\natexlab{b}}.
\newblock
  \urlprefix\url{https://www.caa.co.uk/uk-regulations/aviation-safety/basic-regulation-the-implementing-rules-and-uk-caa-amc-gm-cs/air-traffic-controllers/},
  {A}ccessed: 20-05-2025.

\bibitem[{{CAA}(2020)}]{CAP794}
{CAA}, \emph{{CAP} 794: {Air} {Traffic} {Controllers} – {Initial} {Training}
  {Objectives} - {Civil} {Aviation} {Authority}}, 2020.
\newblock
  \urlprefix\url{https://www.caa.co.uk/our-work/publications/documents/content/cap-794/},
  {A}ccessed: 26-06-2024.

\bibitem[{{ICAO: Montréal, QC, Canada}(2017)}]{icao10056}
{ICAO: Montréal, QC, Canada}, \emph{Manual on Air Traffic Controller
  Competency-Based Training and Assessment}, 2017.

\bibitem[{{The Alan Turing Institute}(2025)}]{bluebird}
{The Alan Turing Institute}, \emph{Project {Bluebird}: {An} {AI} system for air
  traffic control}, 2025.
\newblock
  \urlprefix\url{https://www.turing.ac.uk/research/research-projects/project-bluebird-ai-system-air-traffic-control},
  {A}ccessed: 26-06-2024.

\bibitem[{Kent et~al.(2026)Kent, {De Ath}, Layton, Hart, Everson, and
  Carvell}]{bluebird_mallard}
Kent, P., {De Ath}, G., Layton, M., Hart, A., Everson, R., and Carvell, B.,
  \emph{A Future Capabilities Agent for Tactical Air Traffic Control}, AIAA,
  2026.

\bibitem[{Hansen(2023)}]{hansen_cma_2023}
Hansen, N., \enquote{The {CMA} {Evolution} {Strategy}: {A} {Tutorial},} arXiv,
  Mar. 2023.
\newblock \doi{10.48550/arXiv.1604.00772},
  \urlprefix\url{http://arxiv.org/abs/1604.00772}, arXiv:1604.00772 [cs].

\bibitem[{{de Raadt} et~al.(2021){de Raadt}, Warrens, Bosker, and
  Kiers}]{deRaadt21Reliability}
{de Raadt}, A., Warrens, M.~J., Bosker, R.~J., and Kiers, H. A.~L., \enquote{A
  Comparison of Reliability Coefficients for Ordinal Rating Scales,}
  \emph{Journal of Classification}, Vol.~38, 2021, pp. 519--543.

\bibitem[{Keane et~al.(2026)Keane, Pepper, Burr, Hodgkin, Gould, Korna, and
  Thomas}]{bluebird_assurance}
Keane, A., Pepper, N., Burr, C., Hodgkin, A., Gould, D., Korna, J., and Thomas,
  M., \emph{A framework for assuring the accuracy and fidelity of an AI-enabled
  digital twin of en route UK airspace}, AIAA, 2026.

\bibitem[{Mohamed et~al.(2026)Mohamed, Carvell, Procter, {De Ath}, and
  Everson}]{bluebird_agent_transparency}
Mohamed, E., Carvell, B., Procter, R., {De Ath}, G., and Everson, R.,
  \emph{Towards Transparent AI Agents for Air Traffic Control}, AIAA, 2026.

\bibitem[{Pepper et~al.(2026{\natexlab{b}})Pepper, Thomas, and
  Conti}]{bluebird_data_assim}
Pepper, N., Thomas, M., and Conti, Z.~X., \emph{Fast Surrogate Models for
  Adaptive Aircraft Trajectory Prediction in En route Airspace}, AIAA,
  2026{\natexlab{b}}.

\end{thebibliography}
